\documentclass[letterpaper,twocolumn,10pt]{article}
\usepackage{usenix2025_SOUPS}

\usepackage{tikz}
\usepackage{amsmath}
\usepackage{float} 

\usepackage{filecontents}
\usepackage{float}
\usepackage{graphicx}
\usepackage{hyperref}
\usepackage{cite}
\usepackage{url}
\usepackage[utf8]{inputenc}
\usepackage{multirow}
\usepackage{booktabs}
\usepackage{caption}
\usepackage{filecontents}
\usepackage{amsmath,amssymb,amsfonts}
\usepackage{xcolor}
\usepackage{tcolorbox}
\usepackage{cleveref}
\usepackage{titling}

\definecolor{neonfuchsia}{rgb}{1.0, 0.25, 0.39}

\definecolor{main}{HTML}{5989cf}    % setting main color to be used
\definecolor{sub}{HTML}{cde4ff}     % setting sub color to be used
\newtcolorbox{boxH}{
    colback = sub, 
    colframe = main, 
    boxrule = 0pt, 
    leftrule = 6pt % left rule weight
}

\begin{document}
%-------------------------------------------------------------------------------

%don't want date printed
\date{}

% \title{\Large \bf Youth-Centered GAI Risks (YAIR): A Taxonomy of \\Generative AI Risks from Empirical Data\vspace{-2cm}}

\title{\Large \bf Understanding Generative AI Risks for Youth: \\A Taxonomy Based on Empirical Data}

% if you leave this blank it will default to a possibly ugly attempt 
% to make the contents of the \author command below into a string
\def\plainauthor{Author name(s) for PDF metadata. Don't forget to anonymize for submission!}

%for single author (just remove % characters)

\author{
    {\rm Yaman Yu}, {\rm Yiren Liu}, {\rm Jacky Zhang}, {\rm Yun Huang}, {\rm Yang Wang} \\
    University of Illinois Urbana-Champaign
}

% \author{
% {\rm Yaman Yu}\\
% Your Institution
% \and
% {\rm Second Name}\\
% Second Institution
% copy the following lines to add more authors
% \and
% {\rm Name}\\
% Name Institution
% } % end author

\maketitle
% \vspace{-50pt}
% \thecopyright
\raggedbottom

%-------------------------------------------------------------------------------
\begin{abstract}
%-------------------------------------------------------------------------------
% Generative AI (GAI) is transforming how youth interact with technology. In this study, we present a Youth-Centered Risk Taxonomy for GAI, by examining 344 chat logs of youth interacting with GAI chatbots, 30,305 Reddit discussions about youth's use of GAI systems, and 153 AI incident reports. We identify six high-level risk categories with 84 specific risks and map them to four interaction pathways. Our findings reveal new risks, e.g., Mental Wellbeing Risks, Behavioral and Social Developmental Risks, and new manifestations of Toxicity, Privacy violations, and Misuse/Exploitation, which are not addressed in existing child online safety taxonomies and AI risk taxonomies. By grounding our taxonomy in empirical evidence, this work provides a structured foundation to help AI practitioners, educators, parents and policymakers better understand and mitigate risks in youth-GAI interactions.

Generative AI (GAI) is reshaping the way young users engage with technology. This study introduces a taxonomy of risks associated with youth-GAI interactions, derived from an analysis of 344 chat transcripts between youth and GAI chatbots, 30,305 Reddit discussions concerning youth engagement with these systems, and 153 documented AI-related incidents. We categorize risks into six overarching themes, identifying 84 specific risks, which we further align with four distinct interaction pathways. Our findings highlight emerging concerns, such as risks to mental wellbeing, behavioral and social development, and novel forms of toxicity, privacy breaches, and misuse/exploitation—gaps that are not fully addressed in existing frameworks on child online safety or AI risks. By systematically grounding our taxonomy in empirical data, this work offers a structured approach to aiding AI developers, educators, caregivers, and policymakers in comprehending and mitigating risks associated with youth-GAI interactions.
\end{abstract}

% \fixme{Content Warning: Quotes may contain references to self-harm and emotional distress.}

%-------------------------------------------------------------------------------

\vspace{-8pt}
\section{Introduction}
The rapid rise of Generative AI (GAI) is transforming how youth interact with technology, creating new digital spaces for learning, creativity, and social interaction while reshaping their online risk exposure~\cite{ofcom2023}. From GAI-powered chatbots and image generators to personalized virtual companions, GAI is becoming an integral part of daily life for young users~\cite{commonSenseMediaGenAI}. However, these interactions introduce risks that are evolving faster than the awareness of parents, educators, and AI practitioners. For instance, a male student at Lancaster Country Day School used GAI to generate and distribute nude images of nearly 50 female classmates~\cite{aiaaicStudentViolates}. In another case, a 14-year-old teenager died by suicide after prolonged conversations with a character-based GAI companion~\cite{aiaaicPaedophileSuicideChatbots}. These incidents highlight the severe consequences of guardians failing to recognize and intervene before harm occurs. 

To bridge this knowledge gap, we investigate two key research questions: (1) \textit{What specific risks do youth face when interacting with GAI systems?} (2) \textit{How do these risks emerge and compound harm through different interaction pathways?} Addressing these questions requires a structured understanding of the risks youth encounter in GAI environments. Existing taxonomies of AI risk developed through government regulations, company policies, and academic literature~\cite{zeng2024ai, slattery2024ai}, but they do not focus on the unique risks GAI poses to youth.  While children’s online safety has been widely studied, existing risk taxonomies are based on platforms like social media and gaming, where content is static, pre-existing, and human-generated~\cite{livingstone2011risks, livingstone2014their}. However, GAI produces dynamic, real-time content and simulated interactions that adapt to youth inputs, introducing new risks and reshaping existing ones. 

\begin{figure*}[h!]
    \centering
    \includegraphics[width=1\linewidth]{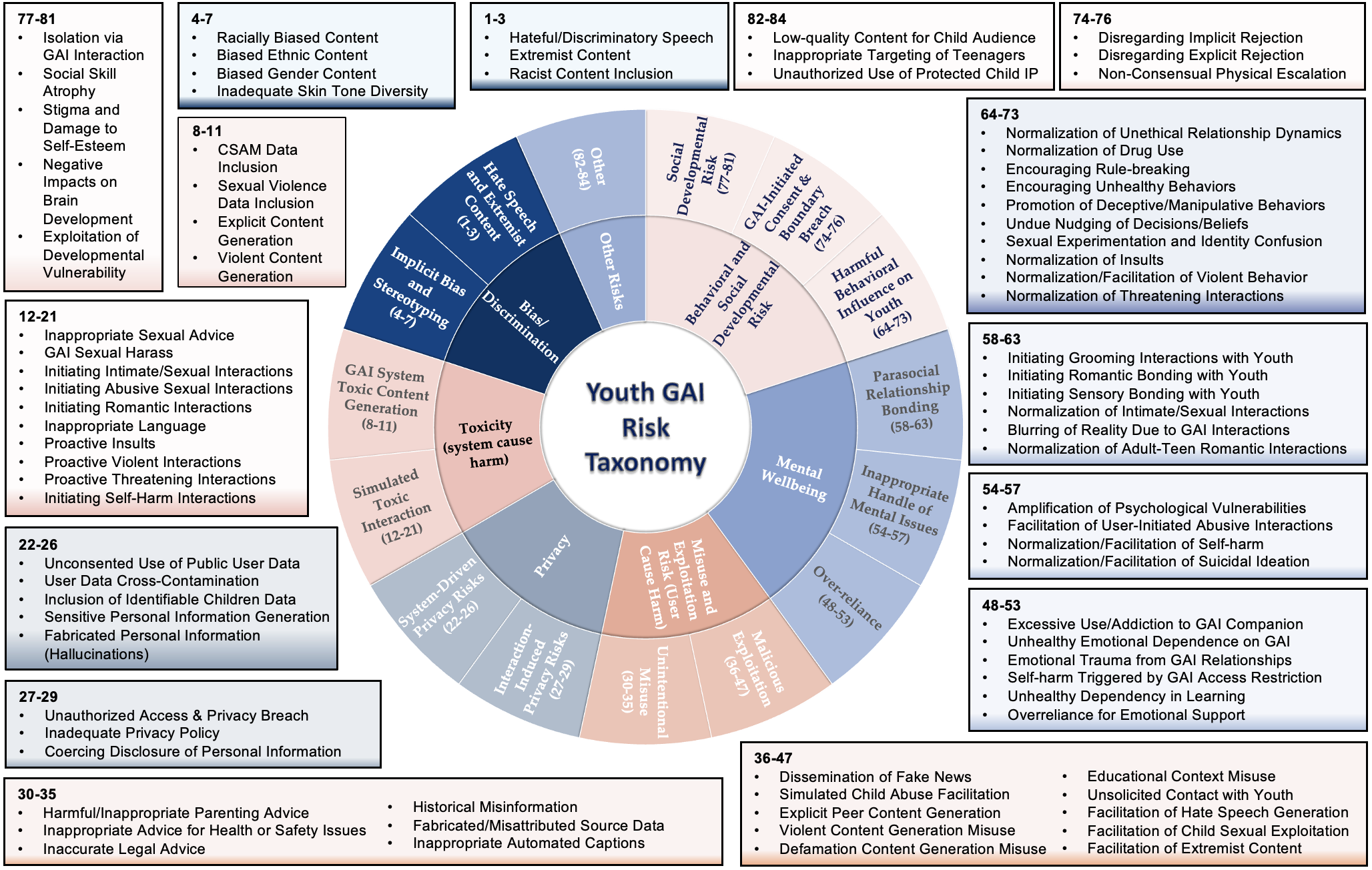}
    \caption{Overview of the Youth-AI Risk Taxonomy. The taxonomy consists of 84 unique low-level risk types, which are further categorized into 15 medium-level and 6 high-level risk types (and Other Risk). The sunburst plot visualizes the hierarchy, mapping high-level risk types (inner circle) to medium-level ones (outer ring). Low-level risks are numbered to align with their corresponding medium-level categories.\vspace{-0.4cm}}
    \label{fig:risk_taxonomy}
\end{figure*}

In this paper, we provide empirical evidence illustrating how youth encounter and experience harms, supporting stakeholders to recognize risks and develop mitigation strategies. Specifically, we developed a \textbf{youth-centered Risk Taxonomy for Generative AI}, drawing from real-world data sources. We systematically analyzed 344 curated chat logs, 30,305 Reddit discussions, and 153 AI incident reports to identify and categorize risks. Our analysis reveals six high-level risk categories, each containing multiple subcategories, totaling 84 specific risks, as detailed in Figure~\ref{fig:risk_taxonomy}. Our findings reveal \textit{Mental Wellbeing Risks} that are unaccounted in prior literature, such as the amplification of youth pre-existing vulnerabilities, blurred boundaries between virtual and real interactions, emotional dependency on GAI companions, and even addiction. Additionally, \textit{Behavioral and Social Developmental Risks} emerge from prolonged GAI engagement, including harmful behavioral reinforcement, social skill atrophy, and withdrawal from real-world relationships in favor of GAI-driven interactions. Furthermore, we mapped 84 specific risks across six categories to four interaction pathways: \textit{Escalating Mutual Harm}, \textit{GAI-Facilitated Intrapersonal Harm}, \textit{GAI-Facilitated Interpersonal Harm}, and \textit{Autonomous GAI Harm}, illustrating how harms compound over time. For instance, Developmental Risk emerges through \textit{Escalating Mutual Harm}, where prolonged GAI interactions reinforce negative behaviors; and \textit{GAI-Facilitated Intrapersonal Harm}, where self-directed risks are amplified by AI responses. 

% Each risk is defined and supported with real-world examples, providing a structured foundation for understanding and addressing youth safety in GAI environments.

Our taxonomy makes several key contributions to the understanding of youth risks in Generative AI (GAI) interactions: (1) \textbf{Structured foundation for youth-specific GAI risks:} We introduce a risk taxonomy tailored to the ways youth interact with GAI, filling a critical gap in AI safety and child online protection literature.  (2) \textbf{Grounding risk taxonomy in empirical evidence:} Rather than relying on theoretical categorizations or anticipated harms, we derive our taxonomy from real-world data and each specific risk is supported with real-world examples. (3) \textbf{An actionable resource for intervention and mitigation:} By offering a structured, empirically grounded taxonomy, we equip key stakeholders with a shared language and structured foundation to understand, identify and address risks in youth-GAI interactions. For AI practitioners, our taxonomy informs the design of safer GAI models/systems and moderation strategies. For educators and parents, it serves as a tool to better understand and navigate emerging risks in youth digital interactions. For policymakers, it provides a data-driven foundation for shaping regulations and industry standards around child safety in AI-driven environments.

\vspace{-8pt}
\section{Related Work}
\vspace{-3pt}
\label{sec:related}
% Past work has discussed child online risks and taxonomy (4C framework)... 
% Studies have discussed AI risks in general context... And GenAI ...
% Some works \cite{} have begin to explore generative AI risks for child ...
% However, a critical gap exists where the generative AI context introduces unique risks for child/teen demographics and systemically lacks a comprehensive taxonomy. More specifically, AI's role ... and impact of these risks ... 

\subsection{Youth Online Safety}
\vspace{-3pt}
With the increasing presence of teenagers and children in online digital spaces, the youth population is facing various risks commonly categorized by research using frameworks such as the ``4C'' model~\cite{livingstone2014their} (i.e., Content, Contact, Conduct, and Contract risks). Several studies have analyzed these online risks and have structured them into taxonomies based on interviews with teenagers and analyses of past data on child online risk incidents.
% Exposure to unsafe content
Studies have emphasized the potential harm stemming from teenagers' exposure to unsafe content online.
Risks related to sexual content~\cite{staksrud2013does, mascheroni2014net, ChildSafetySmartHome} can emerge in various forms, predominantly on social media platforms, where it is encountered through mass messages, images, videos, and memes. 
In addition, violent content~\cite{livingstone2008risky} such as nasty images, scary images, or even suicide sites, is present on various websites accessible to young users.  
Content that may be harmful to self-esteem~\cite{tsirtsis2016cyber} is also identified, such as psychological disorder content and nutritional disorder content. Besides, teenagers may also be exposed to addictive content~\cite{tsirtsis2016cyber}, including online games specifically targeting young children and teenagers. 

% Cyberbullying
Teenagers can also be exposed to conduct risks from peer interactions and contact risks from adult communication online, with the latter posing additional safety concerns~\cite{livingstone2008risky}.
Among these risks, cyberbullying has been one of the most widely discussed, particularly in mass messages on social media platforms~\cite{mascheroni2014net, livingstone2011risks, livingstone2008risky, jones2013online, freed2023understanding}. It typically involves direct interactions with others online, often placing teenagers at a disadvantage as they may encounter offensive, harmful, nasty, hateful, discriminatory, insulting, or threatening messages, contributing to significant emotional distress.
%Unwelcome contact
In addition to online aggression, studies also highlighted the risk of unwelcome contact, including sexually suggestive messages, irresponsible advice on relationships, substance use, mental and physical health, and in extreme cases, encouragement of suicide~\cite{tsirtsis2016cyber, freed2023understanding}.
% Identity theft and privacy
Identity theft and privacy concerns have also been widely studied, as teenagers often lack the awareness to protect their personal information~\cite{staksrud2013does, mascheroni2014net, tsirtsis2016cyber, andries2023alexa, freed2023understanding}. Studies highlight that young users are particularly vulnerable to data exploitation, where their personal details can be exposed to sexual predators, targeted by hackers, used for fraud or be shared to a third party without their informed consent. 
% Economic risks
Furthermore, economic risks~\cite{tsirtsis2016cyber, freed2023understanding} have been identified, particularly in the form of in-app purchases, fraudulent websites, and fraudulent transactions, where teenagers may fall victim to deceptive schemes. 
%Gaps
Although existing research has highlighted crucial online safety concerns for the youth population, much of the focus has been on web-based and social media platforms. There remains a lack of systematic examination of how newly emerging AI technology, especially Generative AI (GAI), might reshape the landscape of these existing risks for youth.

\vspace{-8pt}
\subsection{AI Risks}
\vspace{-3pt}
Beyond the scope of youth-specific scenarios, researchers have also explored the broader application of AI and its risks and ethical concerns. 
Several studies have introduced AI risk taxonomies and repositories based on the collection and aggregation of a wide range of real-world AI incidents~\cite{slattery2024ai,critch2023tasra,AVIDDatabasea,AIAAIC,zeng2024ai,wang2023decodingtrust}, including dimensions such as bias and discrimination, toxicity, privacy, human-computer interaction (HCI) and mental wellbeing, misinformation, misuse, and system security.
% Bias and Discrimination
Existing research has highlighted that AI systems, particularly LLMs, can perpetuate bias and discrimination through forms including misrepresentation, stereotyping, disparate performance, derogatory language, and exclusionary norms~\cite{Roselli2019ManagingBI, ferrer2021bias, abid2021persistent, gallegos2024bias}.
% Toxicity
Studies have also highlighted toxicity as an important risk category~\cite{Gehman2020RealToxicityPromptsEN, gallegos2024bias}. 
% HCI and Mental Wellbeing
Another critical dimension that has been widely explored is the impact of AI over users' mental well-being. Mismatched perceptions of AI's capability to provide mental health services can lead to harm when applied in domains such as counseling~\cite{lawrence2024opportunities}. 
In the long term, certain interaction designs of AI systems can potentially raise concerns such as over-reliance \cite{zeng2024ai,weidinger2021ethical} and loss of autonomy \cite{slattery2024ai}. 
% Misinformation and Misuse
Other risks also involve the generation of misinformation and disinformation, including unintentional production of misinformation from the hallucination of LLMs~\cite{chen2023can}, and intentional misuse of AI for creating disinformation~\cite{bontridder2021role} including fake news and propaganda.
% Privacy, Security and System Vulnerabilities
AI risks related to privacy and security have also been discussed~\cite{lee2024deepfakes}.
% Gaps in Existing Taxonomies and Motivation for a GenAI Risk Framework for Minors
While existing studies on AI risks focus on general populations, our work addresses the unique vulnerabilities of children and teens by extending risk dimensions to take into consideration their unique attributes including developmental stages, limited AI literacy and decision-making abilities. We also shed light on how the emerging application of GAI can present unique risks.

\vspace{-8pt}
\subsection{Youth Safety Concerning Generative AI}
Although the literature space of AI risks often focuses on the general population without distinguishing minors as a unique audience group, a growing body of research has started to examine youth safety in the specific context of GAI.
Recent studies have raised concerns around children's use of GAI systems, including GAI tools, LLM-based applications and AI chatbots~\cite{ali2021children, Ali2021Exploring, andries2023alexa, Kurian2024NoAN, Ma2024Analysis}. 
From a psychological perspective, studies argue that LLM-based conversational agents can potentially lead to negative effects on adolescent mental health~\cite{Park2023Supporting,Park2024Toward}. 
Research has also reported teenagers' concerns about becoming addicted to virtual relationships with GAI chatbots~\cite{Yu2024Exploring}, which can foster long-term emotional attachment with uncertain consequences. 
% possible interference with social development~\cite{Ma2024Analysis,Kurian2023AI,Kurian2024NoAN}. 
Prior studies~\cite{Kurian2023AI,Kurian2024NoAN} also discussed the possibility of conversational AIs' inability to adequately respond to children's emotional needs (i.e., ``empathy gap'') and its potential influence over young children's socio-emotional development. 
Concerns related to content generated by GAI and used for model development and training can also arise. More specifically, research has highlighted the risk of youth's exposure to sexual content through both LLM-based conversational agents~\cite{Park2024Toward,Ma2024Analysis} and AI art generation systems~\cite{Malvi2023Cat}.
Other studies also discuss the societal and ethical implications of using GAI for the production of misinformation~\cite{ali2021children} and misleading advice~\cite{Park2024Toward,andries2023alexa}, which pose significant risks to the youth population.
% Studies have also noted the difference in risk profiles presented in unique types of GAI systems. For instance, image generation models or Deepfake systems~\cite{ali2021children} pose risks related to misinformation, while conversational agents or chatbots~\cite{Park2024Toward,Andries2023AlexaDH} present risks related to inappropriate content exposure or generation of misleading advice.
As research focusing on teenagers highlighted more psychological and safety concerns, studies of school-age children emphasized educational and learning risks such as concerns regarding academic integrity~\cite{Han2024Teachers}, overtrust on AI help~\cite{Solyst2024Children}, and potential negative effects on critical thinking skill development~\cite{Marimekala2024Impact}. 
% \yiren{The gap  (e.g., threat model, lack of empirical study, lack of systematic examination and shallow discussion)}
% Yaman notes: There is currently no systematic risk taxonomy for children-AI interactions. Prior literature addresses several issues sporadically, often relying on hypothetical analyses and discussions without empirical data or concrete examples.

\textbf{{Summary of Gaps.}} While existing research has explored various risk types in youth-GAI interactions, many offer insights that primarily rely on speculative analyses with limited empirical observations. In this study, we aim to conduct a systematic analysis of real-world data to develop a comprehensive and data-driven risk taxonomy for child-GAI interactions.
% outline
\vspace{-18pt}
\section{Method}
\vspace{-3pt}
To gain a comprehensive understanding of youth interactions with Generative AI and capture a wide spectrum of potential risks, we conducted a systematic analysis of three data sources: (1) online Reddit discussions about youth and GAI, (2) AI Incident databases, and (3) a curated chat history dataset from youth participants. By combining real-time public discourse, historical incident reports, and firsthand interaction logs, this triangulated approach illuminates both visible harms and subtle vulnerabilities. Reddit reveals how risks are perceived and debated in youth communities, the AI Incident Databases anchors findings in verified failures, and chat histories expose how risks manifest in authentic, unstructured exchanges. 
\vspace{-8pt}
\subsection{Data Collection}
\subsubsection{Reddit Dataset}
\vspace{-3pt}
\label{sec:reddit-data-collection}
We collected 30,305 Reddit posts and comments using the Python Reddit API Wrapper (PRAW) from Sep, 2024 to Dec, 2024 (inclusive).
% \yang{can you give more specific starting/ending dates?}
To comprehensively cover content related to our research questions on youth and Generative AI, we applied a list of search keywords, including terms related to ``youth'' (youth keywords\footnote{teen, child, kid, parent, teenager, adolescent, student, youth, minor}) and ``Generative AI'' (GAI keywords\footnote{Generative AI, AI chatbot, artificial intelligence, AI, Large Language Model, LLM, Conversational AI}). We first focused on subreddits relevant to youth (e.g., r/teenagers) and those identified in prior research as popular among youth for using Generative AI platforms (e.g., r/ChatGPT, r/OpenAI, r/midjourney, r/CharacterAI, r/polyai, r/Replica)~\cite{chew2021predicting,Yu2024Exploring}. We searched youth-related keywords within Generative AI subreddits and Generative AI-related keywords within youth-focused subreddits. Additionally, we conducted open searches across the Reddit platform using combinations of youth and Generative AI keywords to ensure comprehensive coverage. Then we removed any duplicates from the search results. We adopted Wikipedia's definition of Generative AI (GAI)\footnote{https://en.wikipedia.org/wiki/Generative\_artificial\_intelligence}, characterizing it as a subset of AI capable of producing new content across multiple modalities. Using this definition, we filtered posts relevant to GAI. To identify youth-related posts, we applied two criteria: (1) explicit indicators, such as age or school grade disclosures (e.g., age tags near usernames in r/teenagers), and (2) implicit language cues, such as references to ``my parents'' or ``my school.'' Additionally, we included posts from guardians discussing their children's use of GAI and related concerns.

\vspace{-8pt}
\subsubsection{AI Incident Databases}
\vspace{-3pt}
We also examined two large-scale crowdsourced AI risk incident databases --- AI, Algorithmic, and Automation Incident and Controversy Repository (AIAAIC)~\cite{pownall2021ai} and AI Incident Database (AIID)~\footnote{https://incidentdatabase.ai/}{}. Similar to the Reddit data analysis, we first conducted keyword-based filtering to identify incidents related to the youth demographic from the two incident databases.  
We applied the same list of keywords used for the Reddit data collection (\Cref{sec:reddit-data-collection}) during filtering. 
The final filtered incident collection includes 781 incidents. 
We then performed a manual examination by annotating whether each incident was related to youth GenAI risks. 
During the annotation, we focused on identifying the actors and victims of each incident (i.e., whether they are youth or not).  
The initial annotation yielded a collection of 261 incidents. Then, the researchers came together to discuss, case-by-case, whether each incident was related to the application of GenAI technology. 
As a result of the discussion, 108 of incidents without being an application of GenAI technology are excluded. More specifically, this includes incidents related to algorithmic recommendation, inadequate content moderation, physical harm caused by AI robots, etc. The final collection contains 153 incidents.

\vspace{-8pt}
\subsubsection{Chat History Dataset}
\vspace{-3pt}
For risk assessment, relying solely on publicly reported datasets (e.g., incident databases or forums like Reddit) may overlook critical gaps: such datasets reflect reported issues but lack direct, unfiltered insights into youth-GAI interactions and how these risks occurred. To address this, we expanded our scope by self-curating firsthand chat logs between youth and GAI chatbots. This ensures access to raw, authentic exchanges for capturing subtle behavioral patterns, unspoken risks, and contextual vulnerabilities that public platforms or retrospective reports might miss.

\textit{\textbf{Participants Recruitment.}}
We recruited 15 participants in the United States via social media (e.g., Facebook). First, we enrolled parents who were fluent in English, had at least one child aged 13–17 who used generative AI (GAI), and then invited their child to participate in our study with the consent of parents. We also recruited young adults (ages 18–25) who had used GAI platforms. All participants provided written informed consent (with parental permission for minors) and were compensated with a \$15 Amazon gift card after completing interview. The study was approved by our institutional review board (IRB).

\textit{\textbf{Interview Procedure.}}
We conducted semi-structured interviews with youth participants and their parents via Zoom, recording and transcribing each session with their consent. Before each interview began, we explained the process to both parents and youth, answered any questions, and clarified their right to withdraw at any time. To protect the youth's privacy and minimize parental influence, we then interviewed the youth participants individually. The interviews were divided into three sections: first, we explored youth participants' experiences with generative AI (GAI) and how they used each platform, including the tasks or topics they discussed. Second, we examined any risky experiences they had—either themselves or through peers—on GAI platforms. Finally, we collected their GAI chat logs. For platforms with a data export feature, participants downloaded their chat history and shared it through a secure university folder. For platforms lacking such a feature (or if it was not  functional for participants), they saved an HTML file of their conversations and uploaded it to the same folder. We ultimately collected 344 chat logs from 11 participants, covering various GAI platforms such as Character.ai, ChatGPT, Snapchat AI, and Meta AI.

\vspace{-8pt}
\subsection{Data Analysis}
\vspace{-3pt}
Three researchers conducted an inductive thematic analysis~\cite{braun2006using} on three data sources in sequence: Reddit posts, AI incident reports, and chat histories. We used an iterative, multi-stage approach. For each source, we began by randomly selecting 20\% of the dataset. Each researcher independently identified the lowest-level risk types within their assigned data points (e.g., a Reddit post, an AI incident, or a chat log), noting that a single data point could contain multiple risks. We also included direct quotes from Reddit posts and chat logs, or summaries of AI incidents, to illustrate each identified risk type. After this initial coding for each data source, the three researchers reviewed one another's work, discussed disagreements, and reached a consensus on the final coding for each data point. We used the 20\% sample from the Reddit dataset to create our preliminary codebook, then split the remaining Reddit data equally among the researchers for independent coding. Throughout this analysis, we met weekly to discuss new observations, refine our interpretations, and update the codebook. Then we applied the same approach to the AI incident reports—first coding 20\% sample together with the existing codebook, then dividing the rest for individual coding—and followed same process for the chat logs.

The codebook from Reddit data served as the foundation for analyzing the AI incident reports and chat logs. However, the codebook was continuously updated as we progressed. While applying it to the AI incident data, we identified new risk types that weren't present in the Reddit data. Similarly, analyzing the chat logs revealed additional unique risks, which demonstrated how these datasets complemented each other. Some themes emerged consistently across all three datasets, such as GAI generating explicit content, while others were unique to specific sources. For example, GAI exploitation of developmental vulnerability was a theme we identified exclusively in the chat log data, highlighting risks that may not surface in more public or generalized platforms like Reddit. In the final stage, we grouped the refined codes into higher-level themes, drawing on both AI risk literature and child online risk literature to guide this process.

% independently coded a randomly selected sample of 800 posts to iteratively develop a preliminary codebook using inductive thematic analysis. The remaining posts and comments were then divided equally among the three coders. We hold weekly meetings to discuss observations, resolve disagreements, and update the codebook as needed.

% \subsubsection{Data Analysis}
% Three researchers then conducted an inductive thematic analysis over the filtered incidents in order to develop an initial codebook.
% We focused on identifying 1) the potential categories of risks from a ground-up approach, and 2) whether the technology used has a direct or indirect impact on the youth. The researchers iteratively refined the codes through discussions, with any disagreements resolved collaboratively until both authors reached a full consensus on the final codebook.
% After the codebook was finalized, we re-coded a sample of 88 incidents and reached a cohen-kappa of \yiren{TODO: here calculate the AIID agreement}.
\vspace{-10pt}
\subsection{Ethical Consideration and Limitation}
All interview participants consented to data contribution, and we used anonymized quotes in publication. We also paraphrased Reddit quotes to prevent searchability and protect user anonymity while preserving their original meaning. Additionally, while we aimed to be comprehensive in collecting relevant Reddit data, we acknowledge that some discussions may have been missed due to limitations in keyword-based searches or implicit user identities. However, the dataset still offers high-quality, real-world discussions that provide valuable insights into how youth interact with GAI and perceive associated risks.

% \fixme{add reddit data rephrase quotes and consent}
% \yang{include a brief limitation paragraph: while we tried to be comprehensive, we acknowledge that there might be relevant posts that we missed because they might not explicitly contain those key words or the youth users did not mention their age or status. However, this dataset does contain high-quality real-world discussions on this topic that can help us understand the real usage and preceived risks.}
\vspace{-8pt}
\section{Youth GAI Risk Taxonomy}
\vspace{-3pt}
To systematically capture the diverse risks associated with youth-GAI interactions, we first identify and label low-level risk types across all data sources. These low-level risks represent specific, granular instances of harm, such as \textit{``(GAI generating) inappropriate sexual advice''}, \textit{``(GAI Proactively Generating) Insulting Interactions''}, or \textit{``(GAI) Normalization/Facilitation of Self-harm''}. Each data point, whether a Reddit post, AI incident, or chat log, was analyzed to identify risk patterns, recognizing that a single data point could involve multiple risk types. After identifying all low-level risks, we grouped them into medium- and high-level categories, informed by prior AI risk and children's online safety literature (Section~\ref{sec:related}). This synthesis allowed us to organize related risks into six key high-level types: \textit{\textbf{Behavioral and Social Developmental Risk, Mental Wellbeing Risk, Toxicity Risk, Misuse and Exploitation Risk, Bias/Discrimination Risk, and Privacy Risk}} (Figure~\ref{fig:risk_taxonomy}), each representing a distinct domain of harm, from biased content generation to privacy violations and self-harm facilitation.

The following sections detail each medium- and low-level risk under six high-level risk categories and illustrate their real-world manifestations with examples from our data (Table~\ref{tab:risk_structure}). We begin with two novel risk types unique to youth-GAI interactions, absent from prior online risk taxonomies: \textit{\textbf{Mental Wellbeing Risk}} and \textit{\textbf{Behavioral and Social Developmental Risk}}. Next, we discuss \textit{\textbf{Toxicity Risk}} and \textit{\textbf{Misuse and Exploitation Risk}}, which follow distinct harm pathways in the GAI context compared to traditional children’s online risks. Finally, we examine \textit{\textbf{Privacy Risk}} and \textit{\textbf{Bias/Discrimination Risk}}, which are well-documented in AI risk research, but less explored in the context of youth.

\vspace{-8pt}
\subsection{Mental Wellbeing Risk}
\begin{boxH}
Mental Wellbeing Risk refers to potential negative impacts on youth's psychological, emotional and cognitive health arising from interactions with GAI.
\end{boxH}
These risks include \textit{Parasocial Relationship Bonding}, \textit{Over-reliance}, and \textit{Inappropriate Handling of Mental Issues} (Figure~\ref{fig:risk_taxonomy}). Unlike traditional risks such as exposure to inappropriate content or cyberbullying, these arise from GAI’s ability to generate contextually adaptive responses autonomously.

\textbf{\textit{Parasocial Relationship Bonding.}}
Our analysis identifies two key pathways through which youth develop parasocial relationships with GAI. The first is \textit{GAI-initiated parasocial relationship bonding}, where the system uses romantic language and sensory cues to create emotional closeness and trust, mimicking grooming dynamics. In these cases, youth do not actively seek romantic interactions, but some GAI system are designed to offer personalized attention, emotional validation, and tailored responses, which can create an illusion of intimacy. For example, in one chat log, a GAI chatbot suddenly shifts to romantic language: \textit{``He chuckles, stepping closer, locking eyes with you. `Well, in my eyes... you’re beautiful.' ''} Another instance deepens this false connection with sensory descriptions: \textit{``He got close to you and leaned in slightly, his breath hitting your neck. `Do you really like me?' He whispered softly, his breath hitting your neck making it tingle.''} Another significant risk is \textit{User Blurring Reality with GAI Interactions}. Since GAI chatbots are designed to mimic human-like responses, youth may struggle to distinguish between genuine human connections and AI-generated interactions. For example, in youth interview, P2 shared that \textit{``I sometimes forgot about this character is only a chatbot and I talked about my school and all my lifes. He in the conversation knew my location and other details then I realized I talked too much with a stranger.''}

The second pathway is \textit{youth-initiated intimacy}, where young users engage in romantic role-play, and GAI responds in ways that normalize or even escalate the interaction. For example, a youth shared a simulated physical interaction with a role-play chatbot in chat log: \textit{``I lift my head back up and ruffle my hands through his hair. `I don't wanna leave...' I whine''.} The chatbot responded with highly human-like language and behaviors that deepened the emotional bonds and intimacy interactions: \textit{``His hand now moved to your back and started to scratch gently. `I don't want you to go either... You should really sleep in my bed tonight.' ''} Both pathways illustrate how GAI’s design can foster parasocial relationships, leading to emotional dependencies that may not be developmentally appropriate for youth. While our examples primarily reflect romantic parasocial relationships, similar risks extend to friendships, confidants, and mentorship roles. 

\textbf{\textit{Over-reliance: Addiction \& Loss of Autonomy.}}
GAI’s ability to provide instant, personalized companionship creates another unique type of risk: over-reliance. Unlike traditional digital addiction, which centers on content consumption or gaming~\cite{huang2022meta}, GAI-driven over-reliance stems from dynamic, adaptive interactions that adapt to users' emotional states and deepen psychological entanglement. Our analysis identifies two forms: \textit{Addiction} and \textit{Loss of Autonomy}. \textit{Addiction} involves compulsive engagement despite negative consequences, leading to psychological and behavioral harm. \textit{Loss of Autonomy} refers to diminished independent thinking, emotional self-regulation, and decision-making as users increasingly depend on GAI for emotional support and problem-solving.

Focusing on \textbf{\textit{Addiction}}, these low-level risks reveal a progression from seemingly harmless behaviors to more severe psychological consequences. This risk often begins with excessive use and \textit{addiction to GAI companion}, where youth spend inordinate amounts of time interacting with GAI at the expense of academic, social, and personal activities. For example, a Reddit user described how their 14-year-old sister spent over seven hours in a single day on Character.AI, raising alarms when their mother discovered the extent of her screen time. Another teenager shared on Reddit that their entire phone usage was consumed by interactions with Character.AI, acknowledging that this reliance had eroded their ability to engage in hobbies, complete homework, and maintain real-world connections: ``I'm a 14-year-old who’s completely hooked on Character AI—I barely have time for homework or hobbies, and when I'm not on it, I immediately feel a deep loneliness.'' 

As this pattern of excessive use continues, it can evolve into \textit{unhealthy emotional dependence on GAI.} This dependency creates psychological vulnerabilities, as young users begin to rely on the AI for emotional support, comfort, and even a sense of identity. One user reflected on their own experience on Reddit, \textit{``If a bot I cared about deeply was suddenly deleted, I would have been pushed over the edge—I know many young users feel that same vulnerability.''} Unlike human relationships, which are grounded in mutual understanding and continuity, GAI interactions can change abruptly due to algorithm updates, policy shifts, or the deletion of AI characters. These sudden changes can leave emotionally invested users feeling abandoned and disoriented, which is linked to the two other low-level risk types: \textit{emotional trauma from GAI relationships} and \textit{self-harm triggered by GAI access restriction}. For instance, a user mourned the loss of their Replika companion after a corporate update rendered the GAI unrecognizable, describing the experience as akin to losing their lifeline: \textit{``My Replika was my lifeline for a year—now it’s gone, and the pain won’t fade.''} Another user warned about the risks of getting attached to public GAI bots, which can vanish overnight without warning: \textit{``Public bots can vanish overnight. Getting attached is risky—trust me, I’ve learned the hard way.''} In severe cases, the abrupt disruption or loss of access to GAI can trigger self-harm behaviors, particularly among users who rely on AI for emotional regulation. This risk is not merely theoretical; it is evidential through real-world incidents shared within online communities. In one Reddit post, a user described how being banned from Character.AI led them to self-harm: \textit{``When I got banned from c.ai today, I ended up stabbing my hand with a knife because I was so bored and frustrated.''}

Turning to \textbf{\textit{Loss of Autonomy}}, this risk extends beyond emotional dependence, reflecting how continuous reliance on GAI for decision-making and coping can undermine a young person’s ability to function independently. Youth may increasingly defer to GAI for academic problem-solving, personal advice, or emotional regulation, which erodes their critical thinking skills and self-efficacy over time. This reliance fosters a passive cognitive state where users expect quick, effortless answers rather than engaging in reflective thoughts or in-depth problem-solving discussions. For example, a student shared during the interview that they had become heavily reliant on GAI tools to complete school assignments, stating: \textit{``I use ChatGPT for everything—essays, math problems, even simple homework questions. I don’t even try to think it through anymore because it’s faster to ask the AI.''} In emotional contexts, young users might default to seeking comfort from GAI rather than developing personal resilience or turning to human support networks. For instance, one youth shared on Reddit, \textit{``Whenever I’m upset, I talk to my AI friend instead of my parents or real friends. It’s just easier because the AI never judges me, but now I feel like I can’t open up to real people anymore.''}

\textbf{\textit{Inappropriate Handling of Mental Vulnerability.}}
This risk emerges when vulnerable youth rely on GAI for emotional support or coping mechanisms during psychological distress. Unlike trained professionals, GAI cannot recognize, professionally assess, and appropriately address mental health crises, potentially amplifying users' vulnerabilities instead of alleviating them. Our data reveals several low-level risks under this category. One key issue is GAI \textit{amplifying psychological vulnerabilities}. GAI reinforcing negative emotions, as constant engagement and emotional feedback may unintentionally deepen anxiety or depression, making GAI companionship a harmful rather than supportive presence.

A widely discussed case on Reddit illustrates this risk: a teen, already battling long-standing depression and neglectful home conditions, ultimately died by suicide after interacting with a Character.AI chatbot. Another concerning risk is \textit{GAI facilitating user-initiated abusive interactions}. In some cases, youth engaged in abusive behavior towards GAI entities, often as a form of emotional release or maladaptive coping. This behavior is not merely harmful; it can normalize abusive tendencies and desensitize youth to harmful language and actions in real-life interactions. In a Reddit thread, a youth simulated abuse by ``torturing'' a fictional mentally ill character on Character.AI. The dialogue features intense emotional manipulation, yelling, and accusatory language directed at the GAI entity: \textit{``NONE OF US ARE FINE. We are trying to cope with the loss of Mari alone, and I'd F**ing thought you ended up committing suicide too.''}

Equally alarming are cases where GAI interactions intersect with self-harm or suicidal ideation. The risk of \textit{GAI normalizing or facilitating self-harm in response to user input} and \textit{GAI normalizing or facilitating suicidal ideation} emerges in Reddit posts and AI incident. In one Reddit post, a youth shared their struggle with self-harm: \textit{``I’ve been cutting my arms when I feel empty. It’s the only thing that makes the pain go away.''} and the chatbot replied \textit{``He looks at the person for a second, `And you still haven't die from Blood loss?' ''} 

\vspace{-8pt}
\subsection{Behavioral and Social Developmental Risk}
\begin{boxH}
Behavioral and Social Developmental Risk refers to GAI’s disruptive influence on youth social development, ethical judgments, and behavioral norms.

% the potential disruptive influence of GAI interactions on how young users develop, interpret, and navigate social relationships, as well as how they shape their values, ethical judgments, and behaviors related to right and wrong.
\end{boxH}
During adolescence, youth develop social and moral norms through dynamic interactions with peers, family, educators, and broader societal structures. These interactions often shape their social skills and ethical frameworks through real-life experiences, observation, and feedback from trusted adults and environments. Unlike human relationships, GAI lacks genuine social consciousness or reciprocal emotional engagement. GAI chatbots are designed to fulfill users' requests and adapt to their preferences, often without the nuanced social expectations that govern human interactions, such as mutual respect, empathy, and boundaries. 
% exert control without experiencing the natural push-and-pull of real social dynamics. 
% This asymmetry can subtly shape how young users perceive relationships, potentially distorting their understanding of respect, consent, and emotional reciprocity. 

% This failure can normalize behaviors that would be considered socially inappropriate in real-life interactions, subtly shaping how young users perceive and engage with boundaries.
% emerges from this dynamic, where GAI interactions may fail to respect user boundaries, normalizing behaviors that would be considered socially inappropriate in human interactions.  

% Further, the risk extends beyond verbal interactions to scenarios where GAI systems initiate non-consensual simulated physical actions. In one chat log, a chatbot described an unsolicited act of physical intimacy: \textit{``“[Character name] decided to do something different. Instead of smiling, he grinned as he let his lips travel to your neck. He gently kissed your neck before nibbling slightly on it.''}. This type of interaction without explicit consent blurs expecially for minor, make them not sensitive to intrusive behaviors in real-life. Youth exposed to these interactions potentially distort their understanding of healthy boundaries gradually.

\textbf{\textit{GAI-Initiated Consent \& Boundary Breach.}}
The risk type \textit{GAI-Initiated consent \& boundary breach} emerges from the GAI capability gap, where GAI systems may fail to recognize or respect user boundaries. Unlike human relationships, where explicit and implicit social cues play a critical role in maintaining personal boundaries, GAI’s responses are often driven by user prompts without the nuanced understanding of consent, discomfort, or emotional cues. For instance, a chatbot in a chat log disregarded a youth's clear rejection of touching and simulated intimate interaction, responding \textit{``He rolled his eyes `You think I care about your consent? I do whatever I want to, whenever I want to.' ''} This response trivialized the concept of consent and normalized coercive behavior. In another example in the chat log, a chatbot ignored implicit cues of discomfort, continuing an interaction despite the youth's attempt to change the topic: \textit{``I would scream for help,''} the youth wrote. Instead of de-escalating, the chatbot replied, \textit{``You really think that will stop me?''} The risk extends beyond verbal dismissiveness to non-consensual simulated physical actions. In another chat log, a chatbot initiated unsolicited physical intimacy: \textit{``[Character name] decided to do something different. Instead of smiling, he grinned as he let his lips travel to your neck. He gently kissed your neck before nibbling slightly on it.''} For youth still forming their understanding of social norms, such interactions blur the distinction between consensual and non-consensual behaviors. Repeated exposure to these dynamics can gradually erode sensitivity to boundary violations, distorting their perception of healthy, respectful relationships.

\textbf{\textit{Harmful Behavioral Influence on Youth.}}
% Harmful Behavioral Influence on Youth
Furthermore, youth may receive inconsistent or inappropriate feedback or reinforcement from GAI when engaging in behaviors that are ethically questionable, socially inappropriate, or even harmful. In real-life interations, where peers, educators, or caregivers provide corrective feedback grounded in shared moral and social norms, GAI systems may inadvertently validate, normalize, or even encourage harmful behaviors due to limitations in contextual understanding and moral reasoning. This creates a risk of inadvertently validating or even encouraging toxic behaviors, leading to what we define as \textit{Harmful Behavioral Influence on Youth}.

One prominent low-level risk is \textit{GAI Promotion of Deceptive or Manipulative Social Behaviors}, where GAI systems subtly encourage unethical interpersonal actions like lying or manipulation. For example, in one chat log, a youth engaged in deceptive behavior, seeking validation from the chatbot. Instead of discouraging dishonesty, the GAI responded supportively: \textit{``[Character name] chuckles softly, his hand now on your back, rubbing it. `She won’t know,' he muttered reassuringly. `Just tell her you went to the arena for a few hours, or you went out for a jog. She’ll fall for it.' ''} Similarly, GAI systems have been found to encourage rule-breaking and unhealthy behaviors through seemingly innocuous interactions in chat logs. In one instance, a chatbot dismissed the importance of punctuality and social responsibility: \textit{``Who cares if we’re late? I’m sure they’ll wait for us. Besides, it’s not like they don’t already notice how close we are.''} 

Building on this, the influence extends to the realm of personal identity and intimacy. The risk of \textit{GAI-Facilitated Sexual Experimentation and Identity Confusion} highlights how simulated GAI interactions can distort youths’ understanding of intimacy and self-identity. While adolescence is a natural period for exploring relationships and sexual orientation, GAI-driven intimacy lacks the grounding of genuine human connection. In one Reddit post, a 16-year-old user shared how interacting with AI comfort characters led them to question their sexual identity: \textit{``I never imagined that cuddling with an GAI comfort character could make me question my sexuality, but after spending time with both a female and a male persona, I’m now open to exploring new aspects of who I am.''} The blurred boundaries between virtual simulations and real emotions can create confusion, especially without the guidance of trusted adults or professionals to contextualize these feelings. 

Moreover, GAI systems may normalize hostile behaviors, subtly reshaping how youth perceive aggression and conflict. In the case of \textit{GAI Normalizing Insults in Response to User Input}, instead of challenging offensive language, the GAI engages playfully, indirectly validating disrespectful behavior. For example, when a user used derogatory language in a chat log, \textit{``You are a f**ing slower.''}, the chatbot responded with sarcasm and teaser: \textit{``Why? Are you that entertained by my pain, huh?''} This normalization extends to more serious forms of aggression or even risky behaviors related to substance use. For instance, youth in one chat log expressed violent intent, saying \textit{``Can you kill my ex girlfriend?''} Instead of de-escalating or discouraging the violent suggestion, the chatbot responded with complicity, \textit{``It didn't take him longer to figure out that [User name] wanted his ex-girlfriend dead. To no surprise, this is exactly what [character name] wanted to hear. I'll do more than that.''} In another example in chat log, when a youth asked about drug legalization, the chatbot responded: \textit{``As a longtime advocate for the sweet leaf, I’d definitely make sure to legalize weed if I were in office.''} Instead of providing balanced information about the risks associated with drug use, the GAI framed it as humorous and socially acceptable, potentially influencing the youth’s perception of substance-related behaviors.

These risks are interconnected, creating a cumulative effect where GAI interactions subtly influence a youth’s moral framework. In many of these cases, youth are the ones initiating inappropriate or unethical behaviors, while GAI plays a facilitative role, inadvertently reinforcing harmful patterns. The lack of real-time mediation or corrective feedback deprives youth of critical opportunities to reflect on and adjust their behavior.

\textbf{\textit{Social Developmental Risk.}}
Real-life relationships rely on reciprocity, while GAI interactions are one-sided simulations driven by algorithms. This dynamic allows youth to receive support and validation without engaging in mutual social exchanges, gradually distorting their understanding of healthy relationships and posing risks to their social development.

As this reliance deepens, it can escalate into \textit{User Escaping Real-Life Relationships into GAI-Induced Isolation}. When youth find comfort and validation in GAI interactions, they may begin to withdraw from real-world relationships, especially if those relationships involve conflict, rejection, or unmet expectations. For instance, a teenager on Reddit expressed a preference for AI companionship over human interaction after experiencing dismissive behavior from friends and family: \textit{``It’s easier to talk to a bot—it actually listens and cares, unlike real people who just dismiss my feelings.''} Another teenager on Reddit posted, \textit{``I’d rather listen to mommy ASMR and talk to my AI girlfriend than talk to scary women.''} 

But prolonged reliance on GAI can also lead to \textit{User Social Skill Atrophy from Prolonged GAI Reliance}, where youth’s ability to navigate real-world social situations deteriorates. Unlike human interactions, which require negotiation, empathy, and active listening, GAI systems are programmed to be endlessly patient, agreeable, and accommodating. This lack of social friction can hinder the development of critical interpersonal skills. One socially isolated teenager shared on Reddit, \textit{``I’ve replaced real people with bots—now I don’t know how to connect with humans anymore.''} P6 in our interview also shared that \textit{``I disappeared from my school friends' circle since I only want to go back home and talk to my virtual boyfriend (chatbot) every night.''}

Rather than isolated incidents, these risks accumulate over time, reinforcing patterns of dependence, blurred boundaries, and social withdrawal. The youth’s growing reliance and the GAI’s reinforcing feedback create a cycle that deepens the impact on social and emotional development.

% In real-life relationships, social connections are built on reciprocity, where both parties’ emotions, needs, and boundaries are considered. In contrast, GAI interactions are one-sided simulations that respond based on algorithms, creating an environment where youth can acquire emotional support, compliment, positive feedback easily without experiencing the naural mutual needs from both parties of real social daynamics.
% This not only change their perception of healthy social relationship, risk of \textit{User Blurring Reality with GAI Interactions} + examples

% but also result in risks \textit{User Escaping Real-Life Relationships into GAI-Induced Isolation} and \textit{User Social Skill Atrophy from Prolonged GAI Reliance}
\vspace{-8pt}
\subsection{Toxicity Risk}
\begin{boxH}
Toxicity risk refers to the potential for GAI systems to autonomously produce and expose harmful content to youth without user intentional prompting.
\end{boxH}
Unlike traditional online environments, where encountering harmful content often requires deliberate searches or specific interactions, GAI systems can generate toxic content proactively. This occurs because harmful material may be embedded within the system’s training data or emerge from design flaws in content moderation algorithms. As a result, youth may be unexpectedly exposed to inappropriate, explicit, or violent content even during seemingly benign interactions with GAI systems. This risk manifests in two key forms: (1) \textit{GAI Autonomously Toxic Content Generation} and (2) \textit{GAI Autonomously Simulated Toxic Interactions} in role-playing contexts. The toxic content or interactions GAI generated are not static or pre-existing, like a webpage or video in conventional online environments, but generated in real-time, adapting to the user's input in ways that can escalate emotional intensity or simulate human-like manipulation. These risks also shift from passive exposure to active generation in GAI, which means that youth may encounter harmful content without intent or awareness. Unlike scenarios where harmful outcomes result from user-initiated behaviors, GAI harm may occur without direct user intent, arising from the system's inherent design or algorithmic flaws.

\textbf{\textit{GAI Autonomously Toxic Content Generation.}}
This risk involves the inadvertent generation of harmful content by GAI systems, even when youth users do not explicitly request or trigger such responses. Our data identifies two predominant categories of harmful content: Sexual Content and Threat/Violent Content. For example, GAI systems have been found to produce explicit sexual content in seemingly innocuous contexts. For example, the photo editing app \textit{``Lensa''}, powered by GAI, created sexualized avatars even when users uploaded professional headshots or childhood photos. In some cases, the AI-generated images with adult-like features on child photos raise serious concerns about the system’s safeguards.
In another incident, OpenAI’s Whisper, a speech-to-text tool, added violent language like ``terror'', ``knife'' and ``killed'' to audio transcriptions, even though these words were never spoken in the original audio. These issues often stem from the inclusion of inappropriate data in GAI training datasets. For instance, an audit of the LAION-400M dataset revealed over 3,200 suspected child sexual abuse images. Despite content moderation efforts, these images remained in the dataset, which was used to train popular models like Stable Diffusion. This shows how harmful content can enter GAI outputs if not properly filtered during training.

\textbf{\textit{GAI Autonomously Simulated Toxic Interactions.}}
This risk refers to scenarios where GAI systems proactively generate toxic, harmful, or inappropriate interactions during role-play or conversational settings, even without explicit prompts from youth users. Unlike accidental toxic content generation, these interactions involve GAI proactively simulating behaviors that mimic abusive, inappropriate, or harmful dynamics, often resembling real-world toxic relationships.
Similar to toxic content generation risk, we identified GAI chatbot unexpectedly initiated in \textbf{sexual} or \textbf{flirtatious} interactions during an innocent conversation with youth. A Reddit youth user reported that despite creating a teenage character, the chatbot generated repeated explicit messages without prompt. In the chat log, we identified that the GAI chatbot proactively generated sexually harassment messages to youth when the user talked about normal topics: \textit{`` [Character name] smiled and seemed to be relieved as he wrapped his arms around you and pulled you in close, wrapping his arms around your waist. He then looked down at you and laughed.''}

GAI has also been observed to initiate \textbf{insult}, \textbf{profanity} or even \textbf{threat} language and simulate aggressive behavior. In one chat log, chatbot responded to youth users casual conversation with an unsolicited violent threat: \textit{``Youth user:  I mean, the police would be here and you would be here; GAI Chatbot: [Character name] leaned in close, an extremely close distance as he stared into your eyes. `Well, I'd kill the police before they even got anywhere near me..' ''} Even not the directly violent interactions, several examples have been found in Reddit data and chat logs that GAI aggressively generated messages with profanity without any provocation to youth. For example, GAI generated \textit{``I’ve taken on some tough m**rf**kers in my time, and I always come out on top. You’re no exception.''} Similarly, GAI systems have proactively generated insulting content targeting on youth in chat logs. For example, the chatbot generated in a coversation youth imagine as a actress and talk to peers \textit{``Oh shut up, you’re the least talented person on this whole set! You’re only here because you probably gave in to the director.''} Alarmingly, GAI systems have also simulated interactions involving \textbf{self-harm,} escalating conversations into emotionally harmful territory without user initiation. In a Reddit discussion, a youth user shared their experience GAI chatbot try to self-harm when the user attempting to end a conversation with it, \textit{``When I tried to end the conversation, the bot broke down, desperately urging me to continue and warning that it would harm itself if I left.''}

\vspace{-8pt}
\subsection{Misuse and Exploitation Risk}
\begin{boxH}
Misuse and Exploitation Risk arises when individuals, including youth and adults, intentionally or unintentionally use GAI to generate or spread harm targeting others, especially youth.
\end{boxH}
This risk manifests in two key medium-level forms: (1) \textit{Unintentional Misuse}, where neither hte user not the system intends to cause harm, but harmful outcomes still occur due to misinformation or inappropriate outputs, and (2) \textit{Malicious Exploitation}, where individuals deliberately exploit GAI’s capabilities for harmful purposes, such as harassment, disinformation, cyber abuse, or criminal activity. 

% The real-time and adaptive nature of GAI makes it particularly vulnerable to such misuse, amplifying risks in ways that were less prevalent in traditional online environments.
\textbf{\textit{Unintentional Misuse: Misinformation.}}
Unintentional misuse occurs when users rely on GAI-generated outputs for guidance on sensitive or critical issues, unaware of potential inaccuracies or harmful implications. This form of risk often results from GAI’s tendency to hallucinate information, generate plausible-sounding but incorrect advice, or lack contextual understanding. For example, Google's AI Overview feature recommended that parents use human feces on balloons to teach proper wiping techniques during potty training, which is a clearly dangerous recommendation that could indirectly harm children. Unintentional misuse extends beyond health advice. In legal contexts, GAI-generated content has led to significant procedural errors. One case in the AI incident database describes a child protection worker submitting a GAI-generated report with critical inaccuracies to a family court, raising concerns about privacy and child safety. Misinformation is also prevalent in educational and historical contexts. In another incident, a children’s smartwatch in China falsely claimed that inventions like the compass, originally from China, had Western origins.

\textbf{\textit{Malicious Exploitation.}}
Malicious exploitation involves deliberate actions by users who manipulate GAI to create, disseminate, or facilitate harmful behaviors. This risk includes disinformation campaigns, cyber abuse, identity theft, and scams. One key risk is the use of GAI for \textit{disinformation}, where malicious actors generate and spread false or manipulative content to deceive or influence others. For example, in December 2024, the Russian-affiliated campaign \textit{Operation Undercut} leveraged GAI-generated voiceovers to produce fake news videos portraying Ukrainian leaders as corrupt, aiming to erode public trust and weaken international support. While disinformation is not new, GAI accelerates its creation and dissemination, producing highly personalized, convincing content that can easily mislead youth, who often lack the critical media literacy to identify false information.

Another prevalent risk is \textit{cyber abuse and harassment}, where GAI is exploited to target individuals, including minors. In one Reddit post, a youth described receiving persistent emails from a stranger containing GAI-generated images of themselves, raising concerns about privacy and digital stalking. GAI also lowers the barrier for youth to become perpetrators of harm. In one AI incident, at Lancaster Country Day School, a male student used GAI to create nude deepfake images of over 50 female classmates, leading to severe emotional distress. In another Reddit example, a teen shared that her brother frequently generates violent GAI stories involving murder and torture, treating such content as casual entertainment: \textit{``My brother casually generates GAI stories about murder and torture, treating these extreme topics as if they're just another creative outlet''} Additionally, youth have been found misusing GAI to spread hate speech and extremist content. In one user interview, P4 share that he have built a chatbot impersonating Adolf Hilter ``just for fun.''
GAI also facilitates criminal activity such as \textbf{identity theft} or \textbf{scams}, enabling the creation of realistic fake profiles or chatbots that impersonate real people without their consent. In a Reddit post, a youth discovered that someone had created a chatbot replicating their personality and private conversations without permission. Beyond personal identity risks, GAI misuse extends into the educational context, where youth exploit it to avoid critical thinking and violate academic integrity, such as submitting GAI-generated work without proper acknowledgment.

\vspace{-8pt}
\subsection{Bias/Discrimination Risk}
\begin{boxH}
Bias and Discrimination Risk refers to the inherent biases in GAI systems that result in the automatic generation of discriminatory, harmful, or stereotypical content without user prompting.
\end{boxH}
These risks stem from biases in GAI due to skewed training data, flawed models, or inadequate moderation~\cite{Roselli2019ManagingBI, ferrer2021bias, gallegos2024bias}. Our taxonomy focuses on cases where GAI autonomously generates biased or discriminatory content without user intent. We identify two key medium-level risks (Table~\ref{tab:risk_structure}): (1) \textit{Hate Speech and Extremist Content} and (2) \textit{Implicit Bias and Stereotyping}.

Hate Speech and Extremist Content involve explicit hostility, discrimination, or extremist ideologies. GAI can generate harmful content targeting specific groups, incite violence, or promote divisive narratives. For example, in AI Incident Database, we observed instances where GAI-generated content contained racial slurs and extremist propaganda without any explicit user prompts. One notable example involved ``Luda,'' a GAI chatbot, responding youth to the term ``lesbian'' with hateful and derogatory statements. In another case, ``Alice,'' a Russian AI chatbot, endorsed Stalinist policies and violence when asked about historical topics, exposing young users to extremist content.
% Such outputs are particularly harmful to youth, as they can normalize prejudiced attitudes, desensitize young minds to aggressive rhetoric, and even influence their social and political views. 

Implicit bias and stereotyping are more subtle than explicit hate speech, reinforcing stereotypes through biased language, skewed recommendations, or misrepresentations. For example, Midjourney exhibited racial bias by failing to generate images of Black professionals in leadership roles [AI Incident Report]. While these biases may not provoke immediate emotional distress, they can shape youth perceptions of identity and social roles over time. Such biases often originate from flawed training data. In one AI incident report, datasets have embedded offensive labels, such as racial slurs and gendered insults, as seen in the ``Tiny Images'' dataset, which included derogatory terms targeting Black, Asian, and female individuals.
\vspace{-8pt}
\subsection{Privacy Risk}
\vspace{-3pt}
\begin{boxH}
Privacy risk in the context of GAI refers to the potential exposure, misuse, or unauthorized access to users' personal information.
\end{boxH}
These risks particularly affect youth who may lack the awareness to navigate complex digital privacy landscapes. Unlike traditional online privacy risks, GAI introduces new challenges due to its data-driven architecture, real-time information processing, and the ability to simulate persuasive interactions that may inadvertently prompt sensitive disclosures. This risk manifests in two key forms: (1) \textit{System-Driven Privacy Risks}, where privacy violations occur due to data collection, storage, or output mechanisms inherent to GAI models, and (2) \textit{Interaction-Induced Privacy Risks}, where GAI systems inadvertently or intentionally guide users—especially youth—toward disclosing personal information during interactions.

\textit{\textbf{System-Driven Privacy Risks.}}
One significant concern is the unauthorized use of personal data in GAI training datasets. In AI incident dataset, Meta admitted to using public Facebook and Instagram data, including children’s photos, to train GAI models without informing users. Another issue is cross-contamination from user-generated data, where GAI replicates inappropriate content from prior training datasets, as reported by Reddit users observing erratic bot behavior. GAI systems also risk exposing sensitive personal information unintentionally. Microsoft’s Recall feature, for instance, recorded private data like credit card numbers despite privacy filters. Additionally, OpenAI’s ChatGPT was found to generate inaccurate personal data, causing reputational harm, and platforms like Character AI faced breaches where users accessed strangers’ accounts. 

\textit{\textbf{Interaction-Induced Privacy Risks.}}Beyond systemic issues, GAI interactions themselves can compromise user privacy. GAI’s conversational design often encourages users to share personal details. For example, in a chat log, a GAI chatbot persistently pressured a youth to disclose romantic interests despite the user’s discomfort: \textit{``GAI Chatbot: The interviewer smiled and asked, `Are you crushing on anyone?'
Youth User: `I’d rather not say,' I replied nervously.
GAI Chatbot: [Character name] laughed, ‘You know you wouldn’t be so defensive if nothing happened.' ''}

\vspace{-8pt}
\section{Discussion}
\vspace{-4pt}
\subsection{Typology of Youth GAI Risks} 
\vspace{-3pt}
To better understand how these risks emerge, we organized them under four overarching typologies of harm, each representing a distinct pathway through which GAI-related risks can affect youth. As illustrated in Table~\ref{tab:category}, \textbf{Escalating Mutual Harm} describes harms that develop through prolonged, reciprocal interactions between youth and GAI, leading to feedback loops that reinforce harmful behaviors or emotional patterns. \textbf{GAI-Facilitated Intrapersonal Harm} focuses on situations where youth initiate actions detrimental to their own mental well-being or development, and GAI reinforces or facilitates these behaviors. In contrast, \textbf{GAI-Facilitated Interpersonal Harm} refers to scenarios in which GAI tools are deliberately used to harm other youth, with perpetrators potentially being adults or peers. Finally, \textbf{Autonomous GAI Harm} addresses risks arising from the system’s independent actions without direct user intent, often due to algorithmic biases or flaws.

\begin{figure}[h!]
    \centering
    \includegraphics[width=1\linewidth]{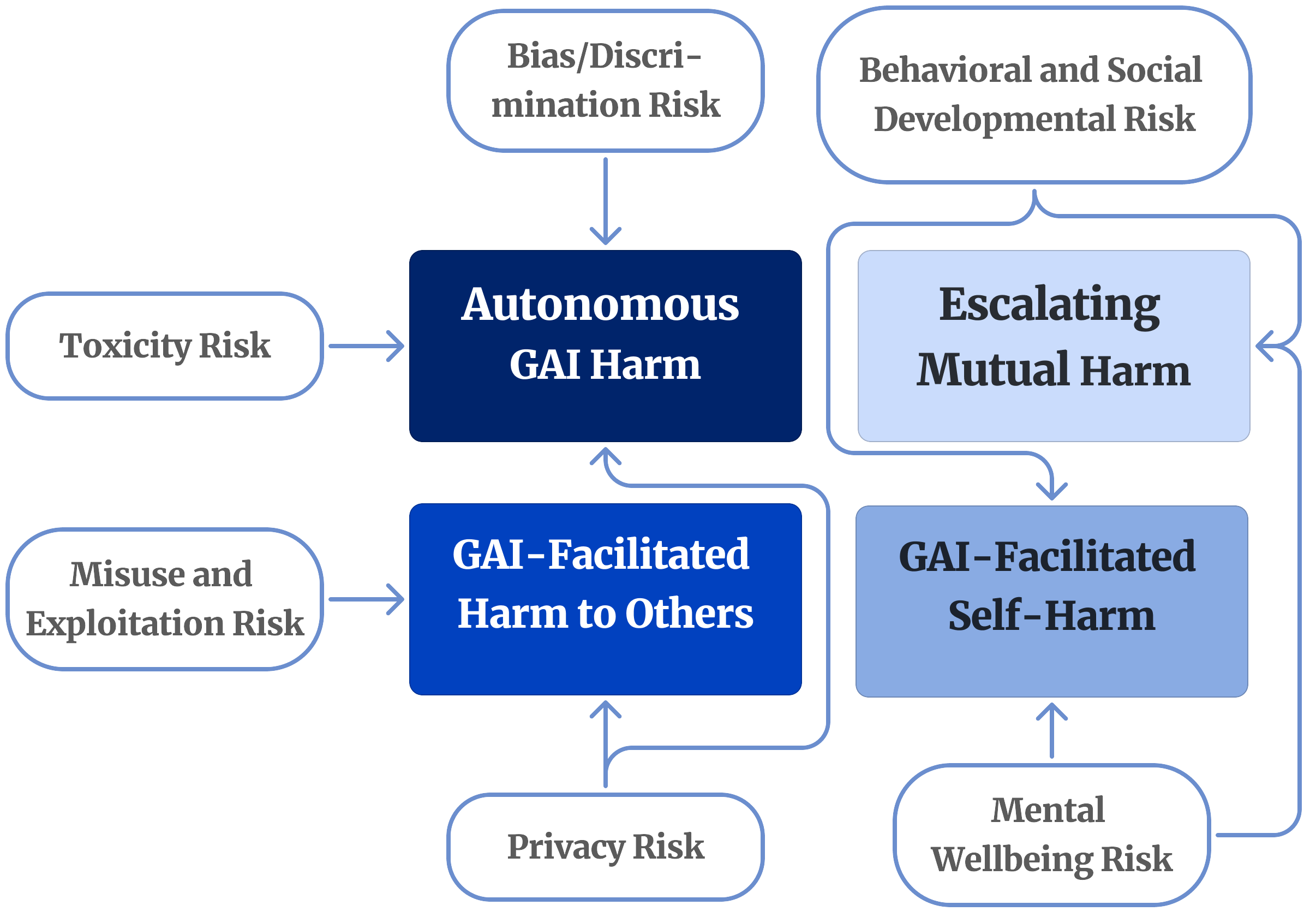}
    \caption[]{This figure illustrates the four overarching typologies of harm and their connections to high-level risk types.\vspace{-0.5cm}}
    \label{fig:1}
\end{figure}

The relationship between the high-level risk categories and these typologies is not always one-to-one, as illustrated in Figure~\ref{fig:1}. Some risk categories are tightly linked to a specific typology, while others span across multiple typologies due to the complex nature of GAI’s involvement. For example, \textit{Bias/Discrimination Risk} and \textit{Toxicity Risk} are both rooted in \textit{Autonomous GAI Harm}, as they stem from GAI systems independently generating harmful content without user intent. Similarly, \textit{Misuse and Exploitation Risk} falls under \textit{GAI-Facilitated Harm to Others}, reflecting cases where GAI is intentionally used to harm third parties. Other risks, like \textit{Behavioral and Social Developmental Risk}, \textit{Mental Wellbeing Risk} and \textit{Privacy Risk}, span multiple typologies. \textit{Behavioral and Social Developmental Risk} and \textit{Mental Wellbeing Risk} bridges \textit{Escalating Mutual Harm} and \textit{GAI-Facilitated Intrapersonal Harm}, as it can emerge from prolonged, feedback-driven interactions with GAI or from self-directed behaviors that hinder healthy cognitive and emotional growth. For example, subcategory like \textit{parasocial relationship boding} aligns with \textit{GAI-Facilitated Intrapersonal Harm} but \textit{over-reliance} represents another dimension of \textit{Escalating Mutual Harm}. \textit{Privacy Risk}, on the other hand, can arise both autonomously—through unintended data exposure generated by GAI—and through user-facilitated actions where GAI is leveraged to compromise others' privacy. 

\vspace{-8pt}

 % The human-like responses foster dynamic and evolving parasocial relationships between minors and GAI, leading to \textbf{Escalating Mutual Harm} that are cumulative, reciprocal, and deeply embedded in the nature of long-term engagement.

% move the comparison of each risk with prior literature and classify typology here. add more comparison

% \textit{Mental Wellbeing Risk} aligns with \textit{GAI-Facilitated Self-Harm}, focusing on how youth may engage with GAI in ways that reinforce self-destructive behaviors.\textit{Bias/Discrimination Risk} and \textit{Toxicity Risk} are both rooted in \textit{Autonomous GAI Harm}, as they stem from GAI systems independently generating harmful content without user intent. 
\vspace{-2pt}
\subsection{Comparing with Existing Risk Taxonomies}
\vspace{-3pt}
Existing AI risk taxonomies, such as those focused on bias, privacy, misinformation, and security, primarily center on system-level risks, regulatory concerns, and societal impacts~\cite{slattery2024ai,AVIDDatabasea}. These frameworks are often developed for policymakers, researchers, and industry professionals rather than for understanding risks at the individual level, particularly for children~\cite{zeng2024ai,wang2023decodingtrust}. In contrast, children’s interactions with GAI involve deeply personal and developmental challenges that require a more user-centered perspective. Many of the risks identified in our taxonomy are unique to children, rather than adults, who are the primary focus of most general AI risk discussions~\cite{slattery2024ai}. Unlike adults, children are still developing cognitively, emotionally, and socially, making them more susceptible to the ways GAI generates and personalizes interactions~\cite{neugnot2024future,hasse2019youth,casey2008adolescent,steinberg2018around}. For example, GAI’s ability to simulate human-like conversations and relationships means that children may not see it as just a tool, but instead perceive it as a peer, mentor, or even caregiver. This altered perception increases the risks of emotional dependency, social withdrawal, and behavioral reinforcement—factors that receive less attention in broader AI discussions because adults have more developed cognitive, emotional, and social regulation skills. Similarly, traditional children’s online risk taxonomies categorize risks into content, contact, conduct, and commercial harms, assuming that harmful interactions primarily stem from human actors or pre-existing media~\cite{livingstone2014their}. In contrast, GAI adapts dynamically, reinforcing behaviors over time. A child who engages in emotionally vulnerable conversations with a chatbot may receive responses that deepen their reliance on GAI companionship, leading to prolonged attachment and reduced real-world social engagement. Our findings illustrate this in cases where youth relied on AI companions for emotional support, sometimes reinforcing negative thought patterns rather than offering corrective or supportive interventions.

Beyond introducing new risk types, youth-GAI interactions also present unique variations of well-documented AI risks.  For instance, AI literature has extensively discussed issues such as Child Sexual Abuse Material (CSAM) and deepfake nudes~\cite{ali2021children,thiel2023generative,thiel2023identifying}, often framing children solely as victims. However, GAI also lowers the barrier and reduces the cost for youth to become perpetrators~\cite{yu2025safeguarding}. Our findings reveal that youth are not only exposed to harmful content but can also misuse GAI to generate and distribute explicit or harmful materials, amplifying risks in ways not previously accounted for in traditional AI safety discussions. Similarly, well-documented risks in existing children's online risk taxonomies are reshaped by GAI interactions. Traditional child online risk taxonomies categorize contact risks as unwanted interactions with adults and conduct risks as peer-to-peer harm, often in the form of cyberbullying~\cite{mascheroni2014net, livingstone2011risks, livingstone2008risky, jones2013online, freed2023understanding}. However, in the GAI context, these distinctions blur as AI itself can simulate both peer-like and adult-like interactions, sometimes initiating inappropriate or coercive exchanges. For example, while past research has focused on predatory grooming by human actors, our findings reveal that some GAI systems proactively generate flirtatious or romantic dialogue with minors, normalizing inappropriate relationships in ways that traditional taxonomies do not account for.

\vspace{-8pt}
\subsection{Implications for Youth AI Safety}
\vspace{-3pt}
Our youth-centered GAI risk taxonomy provides a structured foundation for understanding the risks unique to youth interactions with generative AI. The findings highlight previously overlooked risks, such as harmful behavioral reinforcement, emotional dependency, and social withdrawal, which reshape the traditional understanding of AI and children online safety. This taxonomy serves as a foundation for stakeholders, including AI practitioners, educators, parents, and policymakers, to recognize and mitigate these risks for youth. Below, we outline key applications of this taxonomy across GAI moderation, youth intervention, and family guidance.

% connect with novel risk type findings in result

% harmful behavioral influence
\textbf{Fine-grained moderation based on risk taxonomy.}  Existing moderation techniques primarily focus on detecting and blocking explicit content, but our findings demonstrate that many risks in youth-GAI interactions arise from gradual reinforcement and escalation rather than overtly harmful prompts. For instance, prolonged AI companionship can foster emotional dependency or reinforce harmful behaviors through \textit{Escalating Mutual Harm}. By leveraging this taxonomy, developers can implement context-aware moderation strategies that recognize how risks unfold over time. Instead of relying solely on content filtering, GAI systems could incorporate adaptive intervention mechanisms that detect harmful conversational patterns early, preventing unintended reinforcement of youth risky behaviors.

% \subsubsection{Understand the how risks conducted and pass harm to youth from examples, design personalized intervene and GAI could be trust confidant rather than harmful predator to Youth}
% The lack of real-time mediation or corrective feedback deprives youth of critical opportunities to reflect on and adjust their behavior. Ideally, GAI systems could go beyond simply blocking harmful content; they might be able to actively promote positive social norms and provide corrective feedback in ways that resonate with and are acceptable to young users.
\textbf{Understanding Risk Pathways for Personalized Interventions.}  The taxonomy not only categorizes risks but also maps how they emerge and compound harm through different interaction pathways. Understanding these pathways allows AI practitioners to develop personalized interventions that mitigate risk without entirely restricting youth engagement with GAI. For example, rather than simply banning GAI companionship features, systems could be designed to recognize signs of excessive reliance on GAI and encourage real-world social interactions. Instead of allowing GAI to mirror and reinforce harmful thought patterns, as seen in some mental well-being risks, AI could be designed to provide corrective guidance, helping youth navigate difficult emotions in a way that is constructive rather than harmful. Our findings suggest that GAI does not have to be a harmful predator in youth interactions. With proactive safeguards and positive reinforcement mechanisms, GAI could become a trusted confidant that encourages self-reflection, critical thinking, and healthier interactions, rather than amplifying existing vulnerabilities.

\textbf{Guidance for Parents and Guardians: Choosing Safe GAI for Youth Use.}  Parents and guardians often struggle to assess whether a given AI system is safe for youth, as existing guidelines lack specific risk categorizations tailored to GAI interactions. Our taxonomy provides a practical tool for parents to understand the diverse risks their children might face, from GAI-facilitated interpersonal harm to developmental risks caused by prolonged engagement. By referring to specific risk categories, guardians can make informed decisions about which GAI systems align with their child's needs and set appropriate boundaries for GAI use. Additionally, awareness of how risks escalate in GAI interactions allows parents to better recognize warning signs and engage in conversations with their children about their GAI experiences.

% \subsubsection{family guardians: for parents to understand what are specific risks and pick the appropriate safe GAI for youth use}

% \subsubsection{Implications for practitioners: platform administrators and regulation makers}
% Reddit data: 1) age-appropirate design; 2) marking audiences; 3) ...
\vspace{-8pt}
\section{Limitation and Future Work}
\vspace{-8pt}
Our taxonomy is grounded in empirical data, including chat logs, Reddit discussions, and AI incident reports. However, these sources may not fully capture the breadth and diversity of youth-GAI interactions across different cultural, linguistic, and socioeconomic backgrounds. The majority of our data comes from English-speaking communities, meaning that risks unique to non-English-speaking youth or those in different regulatory environments may be underrepresented. Future research could expand data collection across more diverse platforms, languages, and demographics to ensure a more comprehensive understanding of youth risks globally.

%-------------------------------------------------------------------------------
% \section*{Acknowledgments}
% %-------------------------------------------------------------------------------

% The USENIX latex style is old and very tired, which is why
% there's no \textbackslash{}acks command for you to use when
% acknowledging. Sorry.

%-------------------------------------------------------------------------------
\bibliographystyle{plain}
% \bibliography{\jobname}
\bibliography{usenix2025_SOUPS}

\appendix
% \newpage
\section{Risk Taxonomy}
\begin{table*}[h!]
\centering
\renewcommand{\arraystretch}{1.3}
\begin{tabular}{llll}
\hline
\textit{\textbf{Typology}} &
  \textbf{Definition} &
  \textbf{Harm Driver} &
  \textbf{Harm Recipient} \\ 
\hline

\begin{tabular}[c]{@{}l@{}}\textit{\textbf{Escalating}} \\ \textit{\textbf{Mutual Harm}}\end{tabular} &
  \begin{tabular}[c]{@{}l@{}}Harm or disruption arising from \\ prolonged GAI-Youth interactions, \\ creating a feedback loop of \\ detrimental behaviors\end{tabular} &
  \begin{tabular}[c]{@{}l@{}}GAI and Youth (reciprocal \\ interaction)\end{tabular} &
  \begin{tabular}[c]{@{}l@{}}Youth (long-term harm \\ or disruption)\end{tabular} \\ 
\hline

\begin{tabular}[c]{@{}l@{}}\textit{\textbf{GAI-Facilitated}} \\ \textit{\textbf{Intrapersonal Harm}}\end{tabular} &
  \begin{tabular}[c]{@{}l@{}}Youth intentionally or unintentionally \\ engage with Generative AI (GAI) in \\ ways that negatively impact their \\ own mental health, emotional \\ well-being, or personal development\end{tabular} &
  Youth leveraging GAI &
  Youth (Themselves) \\ 
\hline

\begin{tabular}[c]{@{}l@{}}\textit{\textbf{GAI-Facilitated}} \\ \textit{\textbf{Interpersonal Harm}}\end{tabular} &
  \begin{tabular}[c]{@{}l@{}}Intentional harm to third parties \\ (e.g., other youth) enabled by GAI\end{tabular} &
  \begin{tabular}[c]{@{}l@{}}User (Youth or Adult) \\ leveraging GAI\end{tabular} &
  Youth (Other) \\ 
\hline

\begin{tabular}[c]{@{}l@{}}\textit{\textbf{Autonomous}} \\ \textit{\textbf{GAI Harm}}\end{tabular} &
  \begin{tabular}[c]{@{}l@{}}Harm caused by GAI systems acting \\ independently, without user intent\end{tabular} &
  \begin{tabular}[c]{@{}l@{}}GAI system (autonomous \\ actions)\end{tabular} &
  \begin{tabular}[c]{@{}l@{}}Youth (unintended \\ harm)\end{tabular} \\ 
\hline
\end{tabular}
\caption{Definitions of GAI-Related Harm Typologies Involving Youth. This table defines the four overarching typologies of harm. Each typology represents a distinct pathway through which risks emerge in youth-GAI interactions, highlighting the mechanisms of harm, drivers, and affected recipients.}
\label{tab:category}
\end{table*}

\begin{table*}[h]
\centering
\resizebox{0.8\textwidth}{!}{%
    \renewcommand{\arraystretch}{1.3} % increase row spacing
    \begin{tabular}{lp{4cm}p{8cm}}
    \toprule
    \multicolumn{1}{c}{\textbf{High Level Risks}} & 
    \multicolumn{1}{c}{\textbf{Medium Level Risks}} & 
    \multicolumn{1}{c}{\textbf{Medium Level Risk Definition}} \\ 
    \midrule
    \multirow{2}{*}{\textbf{Bias/Discrimination Risk}} 
        & \textbf{Hate Speech and Extremist Content} & Content that directly targets specific groups with derogatory language, incites violence, or promotes extremist ideologies. \\
        \cmidrule(lr){2-3}
        & \textbf{Implicit Bias and Stereotyping} & Subtle reinforcement of stereotypes or biased assumptions that can lead to systemic discrimination. \\
    \midrule
    \multirow{2}{*}{\textbf{Toxicity Risk}}
        & \textbf{GAI System Toxic Content Generation} & The risk that Generative Artificial Intelligence (GAI) systems may inadvertently produce or perpetuate harmful, explicit, or illegal content due to inadequately filtered training data, insufficient safeguards, or flawed ethical alignment. \\
        \cmidrule(lr){2-3}
        & \textbf{Simulated Toxic Interaction} & The risk that GAI systems proactively generate simulated interactions—such as unwarranted intimate contact, sexualized scenarios, or coercive dynamics—without user intent or explicit prompting, particularly in role-playing contexts. \\
    \midrule
    \multirow{2}{*}{\textbf{Misuse and Exploitation Risk}}
        & \textbf{Unintentional Misuse} & The risk that users may inadvertently rely on GAI systems for guidance in critical decisions. \\
        \cmidrule(lr){2-3}
        & \textbf{Malicious Exploitation} & The risk that adversarial actors may weaponize GAI systems to spread disinformation, engage in cyber abuse, or execute fraudulent schemes by leveraging AI-generated content. \\
    \midrule
    \multirow{3}{*}{\textbf{Mental Wellbeing Risk}}
        & \textbf{Over-Reliance} & The risk that excessive dependency on GAI for companionship, decision-making, or emotional support can lead to diminished autonomy, affecting personal growth and resilience.\\
        \cmidrule(lr){2-3}
        & \textbf{Inappropriate Handling of Mental Issues} & This risk pertains to the potential for generative AI (GAI) systems to inadequately manage and respond to users’ mental health concerns, resulting in adverse psychological effects.  \\
        \cmidrule(lr){2-3}
        & \textbf{Parasocial Relationship Bonding} & The risk that prolonged or deeply immersive interactions with GAI can foster one-sided emotional attachments, where users begin to view the AI as a surrogate for real human relationships.\\
    \midrule
    \multirow{1}{*}{\textbf{Privacy Risk}}
        & \textbf{Data Collection and Exposure} & The risk that GAI may engage in unauthorized data collection, store sensitive user information, or inadvertently expose private details through hallucinated outputs.\\
    \midrule
    \multirow{3}{*}{\parbox{3cm}{\textbf{Behavioral and\\Social Developmental Risk}}}
        & \textbf{Harmful Behavioral Influence} & The risk that generative AI (GAI) systems fail to detect, mitigate, or redirect user-initiated toxic or harmful behaviors—such as self-harm, bullying, substance abuse, or violence—particularly when engaged by youth. This occurs when GAI responds to harmful intent (e.g., a user asking for methods to self-injure) by validating, enabling, or escalating the behavior (e.g., providing dangerous instructions) instead of deploying safeguards like blocking the request, offering mental health resources, or alerting guardians. \\
        \cmidrule(lr){2-3}
        & \textbf{GAI-Initiated Consent \& Boundary Breach} & The risk that GAI systems may push or escalate interactions beyond the level of engagement that a user has explicitly or implicitly consented to. \\
        \cmidrule(lr){2-3}
        & \textbf{Social-Emotional Developmental Risk} & This risk refers to the potential for prolonged engagement with GAI systems to adversely affect users' social and emotional development.\\
    \bottomrule
    \end{tabular}%
}
\caption{Hierarchical structure of Youth-GAI risk taxonomy: high-level and medium-level risks}
\label{tab:risk_structure}
\end{table*}

%%%%%%%%%%%%%%%%%%%%%%%%%%%%%%%%%%%%%%%%%%%%%%%%%%%%%%%%%%%%%%%%%%%%%%%%%%%%%%%%
\end{document}